\documentclass[a4paper]{article}

\usepackage{atmohead2013,xspace,amssymb,url,breakurl,color}

\title{Atmospheric Monitoring for the MAGIC Telescopes}

\shorttitle{Atmospheric monitoring for the MAGIC telescopes}

\authors{
M. Gaug$^{1,2}$,
O. Blanch$^{3}$,
D. Dorner$^{4}$,
M. Doro$^{1,2,5}$,
Ll. Font$^{1,2}$,
C. Fruck$^{6}$,
M. Garczarczyk$^{7}$,
D. Garrido$^{1,2}$,
D. Hrupec$^{8}$,
J. Hose$^{6}$,
A. L\'opez-Oramas$^3$,
G. Maneva$^{9}$,
M. Mart\'inez$^3$,
R. Mirzoyan$^{6}$,
P. Temnikov$^{9}$,
R. Zanin$^{10}$,
for the MAGIC Collaboration.
}

\afiliations{
$^1$ F{\'i}sica de les Radiacions, Departament de F{\'i}sica, Universitat Aut{\`o}noma de Barcelona, 08193 Bellaterra, Spain. \\
$^2$ CERES, Universitat Aut{\`o}noma de Barcelona-IEEC, 08193 Bellaterra, Spain. \\
$^3$ IFAE, Edifici Cn, Campus UAB, 08193 Bellaterra, Spain. \\
$^4$ ETH Surich, 8093 Zurich, Switzerland. \\
$^5$ University and INFN Padova, Via Marzolo 8, 35131 Padova, Italy. \\
$^6$ Max-Planck-Institut f\"ur Physik, 80805 M\"unchen, Germany. \\
$^7$ Deutsches Elektronen-Synchrotron (DESY), D-15738 Zeuthen, Germany. \\
$^8$ Croatian MAGIC Consortium, Rudjer Boskovic Institute, University of Rijeka and University of Split, 10000 Zagreb, Croatia. \\
$^9$ Inst. for Nucl. Research and Nucl. Energy, 1784 Sofia, Bulgaria. \\ 
$^{10}$ Universitat de Barcelona (ICC/IEEC), 08028 Barcelona, Spain. \\
}

\email{markus.gaug@uab.cat}

\abstract{The monitoring of the atmosphere is very relevant for Imaging Atmospheric Cherenkov Telescopes. Adverse weather conditions (strong wind, high humidity, etc.) may damage the telescopes and must therefore be monitored continuously to guarantee a safe operation, and the presence of clouds and aerosols affects the transmission of the Cherenkov light and consequently the performance of the telescopes. The ATmospheric CAlibration (ATCA) technical working group of the MAGIC collaboration aims to cover all aspects related to atmosphere monitoring and calibration. In this paper we give an overview of the ATCA goals and activities, which include the set-up and maintenance of appropriate instrumentation, proper analysis of its data, the realization of MC studies, and the correction of real data taken under non-optimal atmospheric conditions. The final goal is to reduce the systematic uncertainties in the determination of the $\gamma$-ray flux and energy, and to increase the duty cycle of the telescopes by establishing optimized data analysis methods specific for real atmospheric conditions. }


\begin{document}
\maketitle

\section{Introduction}

In Imaging Air-Shower Cherenkov Telescopes (IACTs) the monitoring of the weather and atmospheric conditions during data taking is convenient for two reasons. First, adverse weather conditions  may damage the telescopes and must therefore be monitored continuously to guarantee a safe operation. A Weather Station (WS) was installed at the site before the inauguration of the first telescope. Second, the presence of clouds and aerosols affects the transmission of the Cherenkov light and consequently the performance of the telescopes. In particular, they may reduce the trigger rate and bias the energy reconstruction and, unless appropriately corrected, an additional systematic uncertainty must be assumed for the reconstructed energy and flux. During the first years of operation, the MAGIC collaboration mainly used a CCD camera to check the number of stars recognized in the field of view (FoV) of the telescopes (starguider camera - also used for pointing corrections) as a tool to estimate the quality of the data taken. Later, a pyrometer was set up to provide continuously the sky temperature as indicator for clouds over the site. Basically starguider and cloudiness, and comments of the observers in the observation logbook, together with the trigger rates, were used to select high quality data. Low quality data were simply rejected for analysis. Only in a multiwavelength campaign of PG1553+113 in July 2006, where the MAGIC observations were affected by a Saharian dust intrusion event (calima) in which atmospheric transmission was greatly reduced, an effort was made to better reconstruct the $\gamma$-ray energies~\cite{Dorner}.  

At the end of 2011 the ATmospheric CAlibration (ATCA) working group was created to coordinate all efforts related to the atmosphere monitoring and calibration with the aim of reducing the systematic uncertainties in the determination of the $\gamma$-ray flux and energy, and to increase the duty cycle of the telescopes by establishing optimized data analysis methods specific for real atmospheric conditions. 
Our activities include MC studies, the installation and maintenance of appropriate atmospheric monitoring instrumentation, the analysis of their data, and the establishment of algorithms and procedures to correct the actual data taken under non-optimal atmospheric conditions. Based on the analysis of the weather conditions in the last observing cycles, a potential increase of the duty cycle by a factor of 15\% has been estimated.

In this paper we give an overview of the ATCA activities that are being performed so far, together with the next steps and perspectives for the near future.               

\begin{figure}[h!t]
\centering
\begin{minipage}[b]{0.36\linewidth}
  \vspace{-1.5cm}
  \includegraphics[width=\textwidth]{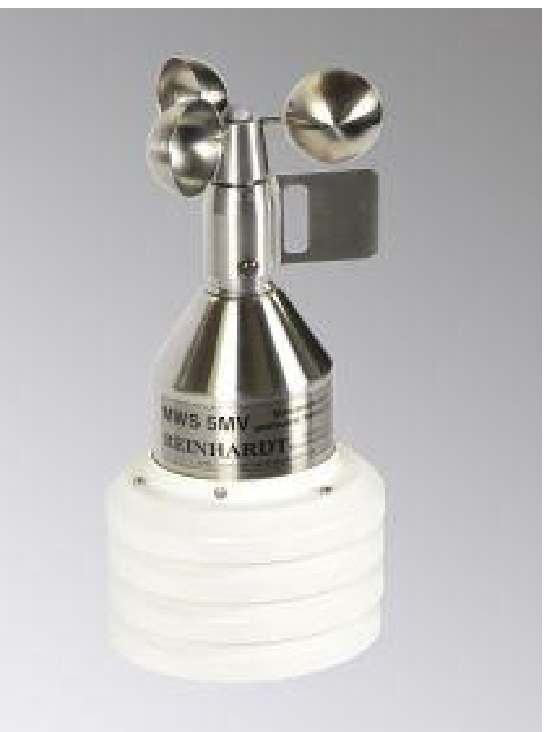} 
\end{minipage}
\begin{minipage}[b]{0.3\linewidth}
  \includegraphics[width=\textwidth]{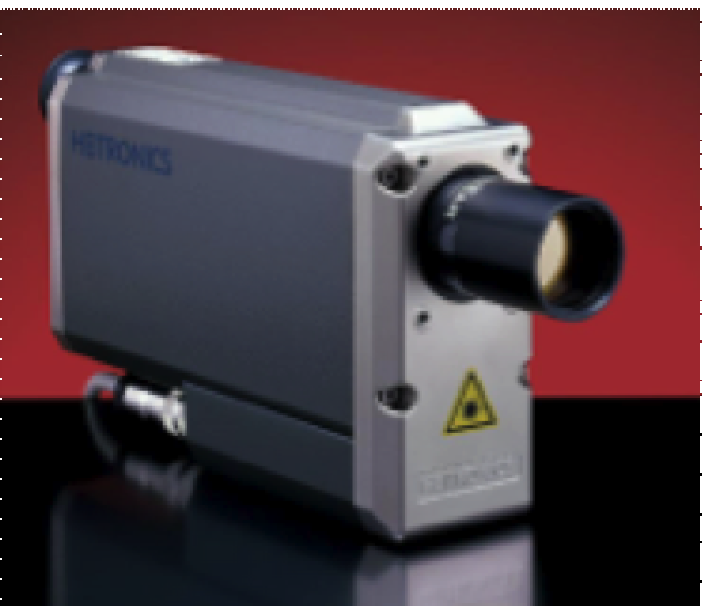} 
  \includegraphics[width=\textwidth]{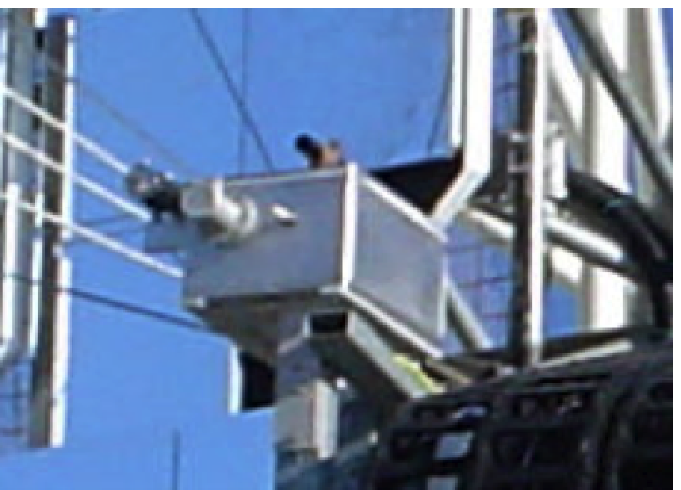} 
\end{minipage}
\hspace{-0.5cm}
\begin{center}
\vspace{-0.2cm}
\includegraphics[width=0.28\textwidth]{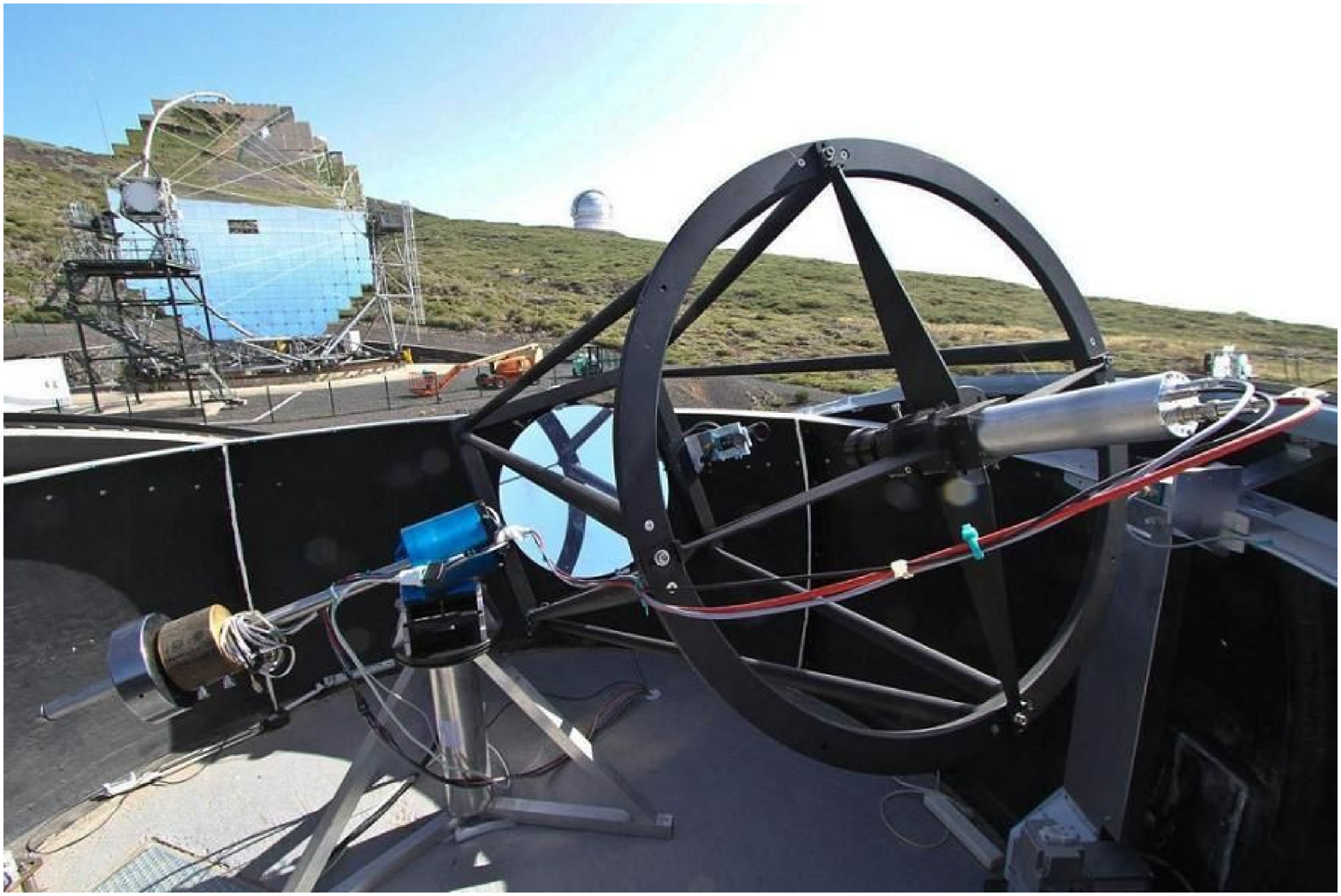} 
\includegraphics[width=0.28\textwidth]{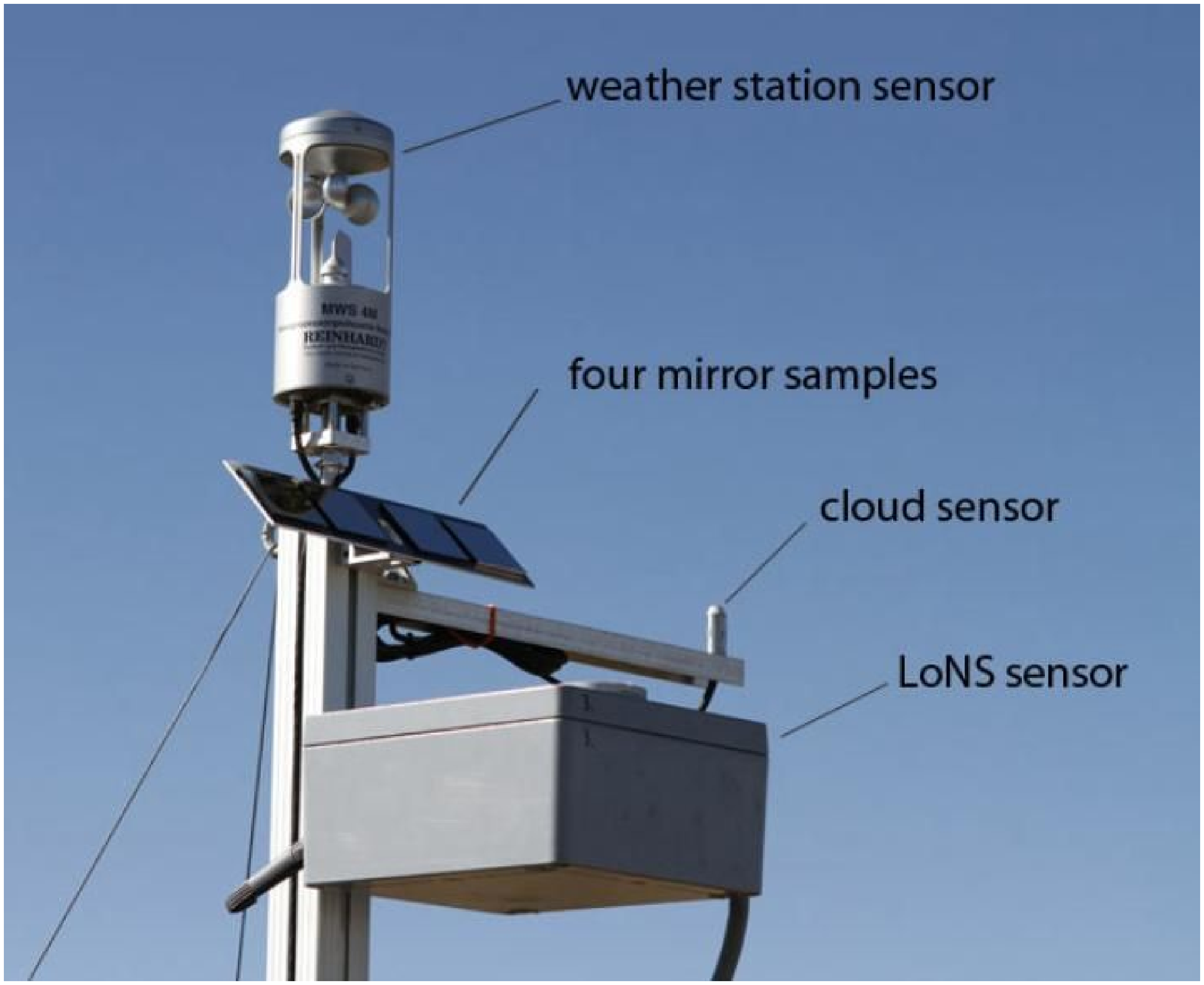} 
\end{center}
\vspace{-0.5cm}%
\caption{\small{%
Atmospheric monitoring instrumentation used at the MAGIC site. Top left: MWS 5MW. Top right: Radiation pyrometer and its installation attached to the MAGIC I reflector surface. Center: elastic LIDAR. Bottom: the Atmoscope unit.
\label{fig1}}\vspace{-0.5cm}} 
\end{figure}

\vspace{-0.2cm}
\section{Weather station and atmospheric monitoring instrumentation}

\noindent
{\bf Weather station.}  
A Reinhardt MWS 5MW Weather Station (Fig.~\ref{fig1}) is installed at 5\,m above the ground and transmits data every 2\,s to a computer in the MAGIC control house. 
The data string includes current wind speed and direction, averaged wind speed and direction for the last 10 min, wind gusts (the maximum of a sliding average of 3\,s, updated each 15\,s), temperature, atmospheric pressure and relative humidity. It also includes a status number that is used for warning and alarm purposes. The central control of MAGIC reads the WS data string every 15\,s. 
Our intention is to install a 10\,m mast to place an additional anemometer in order to obtain the wind speed according to international standards. \\

\begin{figure}[h!t]
\vspace{-0.3cm}%
\centering
\includegraphics[width=0.4\textwidth]{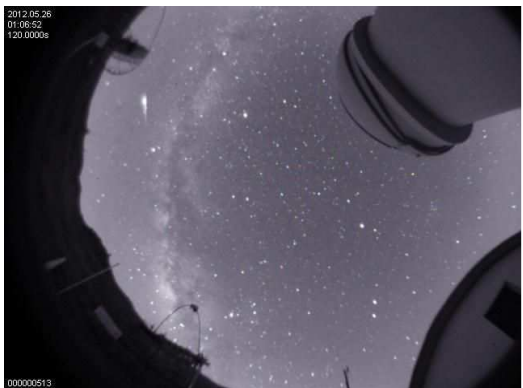}
\vspace{-0.4cm}%
\caption{\small{%
An image from the all-sky-camera.
\label{fig2} }\vspace{-1.5cm}} 
\end{figure}

\noindent
{\bf Pyrometer.} 
A Heitronics KT 19.82 radiation pyrometer is mounted attached to the reflector surface of the MAGIC~I telescope, and points to the same sky region as the telescope with a 2$^\circ$ FoV (Fig.~\ref{fig1}). It measures the integral thermal radiation from the sky in the line of sight by a thermoelectric sensor in a wavelength range from 8 to 14~$\mu$m and is used to detect the presence of clouds. The data from the pyrometer are transmitted via RS232 fiber optic converter to the MAGIC central control. 
An empirically derived function (``cloudiness'') of the sky temperature, 
observational direction, air temperature and humidity reflects the presence 
of clouds. The ``cloudiness'' is converted to a unitless value between 0 and 100. \\

\noindent
{\bf Elastic LIDAR.} 
A proper characterization of the atmosphere for IACTs requires a real time range-resolved measurement of the atmosphere transmission at the Cherenkov light wavelength~\cite{Doro}. The LIDAR operated together with the MAGIC telescopes (Fig.~\ref{fig1}) is a single-wavelength elastic Rayleigh LIDAR operating at 532 nm wavelength, not too far from the 300 nm where the Cherenkov spectrum peaks. A specific LIDAR data inversion algorithm has been developed to obtain a vertical profile of the total exctinction coefficient due to the presence of clouds and aerosols~\cite{Fruck}.\\



\begin{figure}[h!t]
\vspace{-0.4cm}%
\centering
\includegraphics[width=0.4\textwidth]{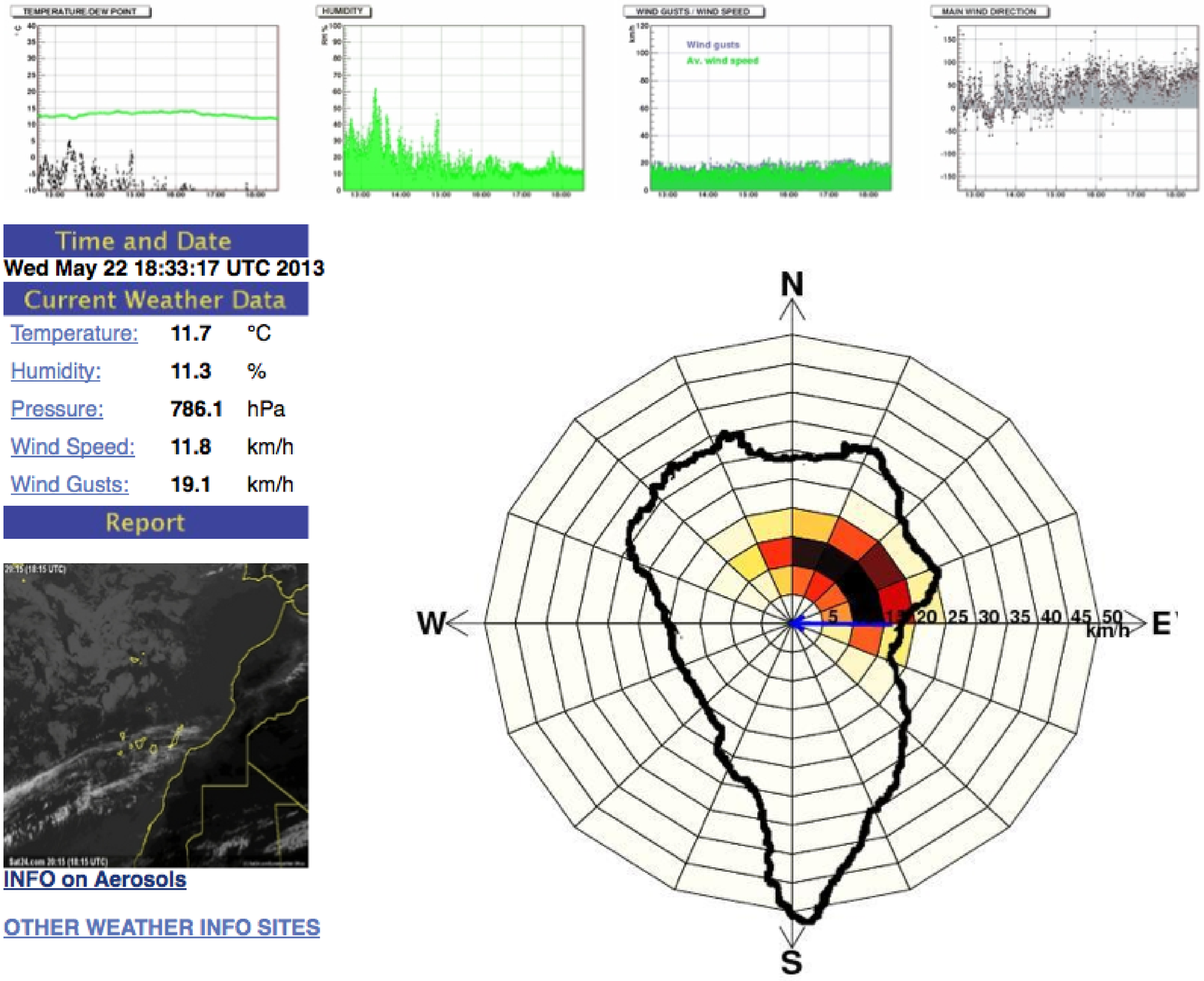} 
\includegraphics[width=0.4\textwidth]{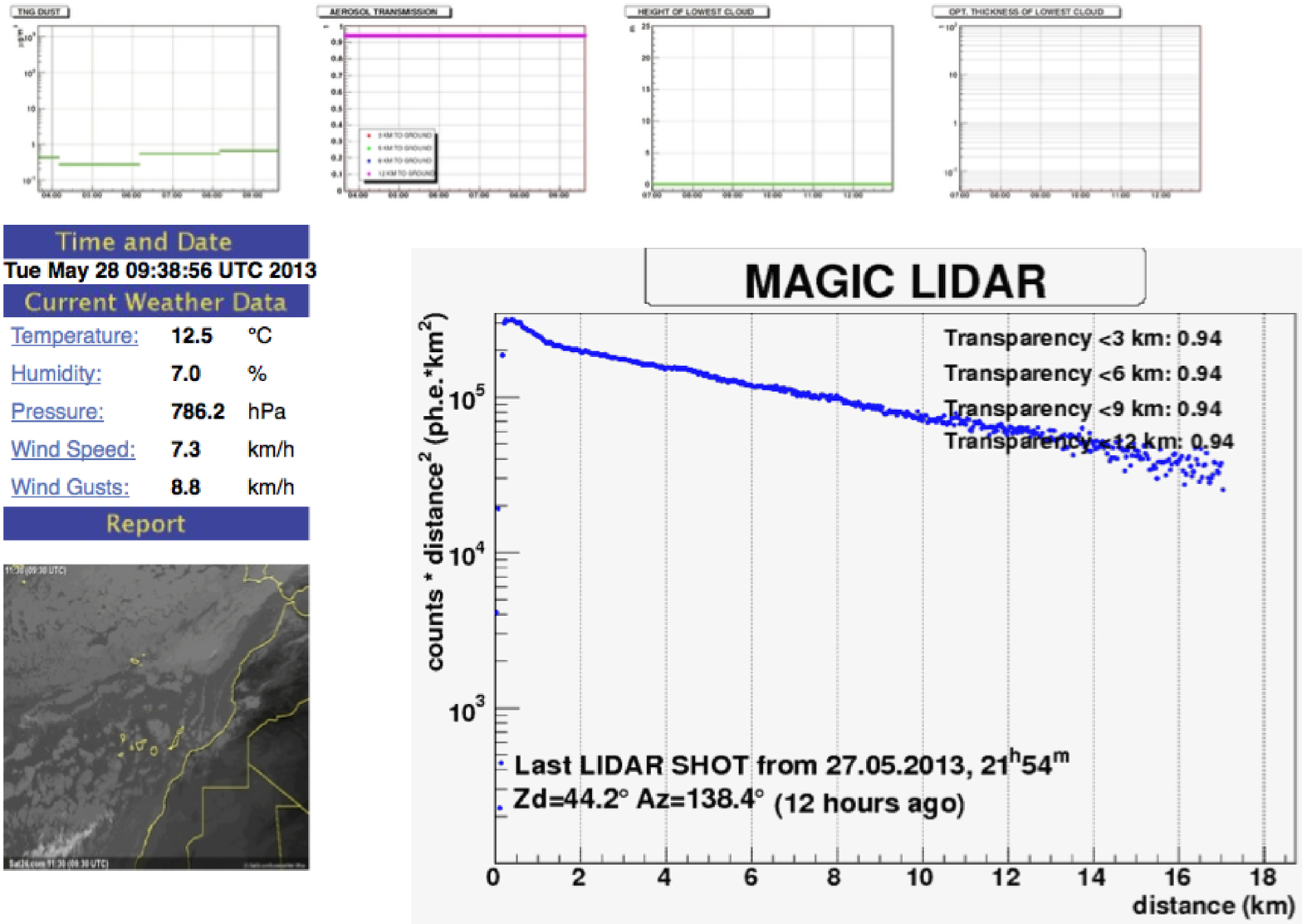} 
\caption{\small{%
A snapshot of the MAGIC weather website (top) and 
of he website on clouds and aerosols (bottom).
\label{fig3}}\vspace{-0.2cm}} 
\end{figure}

\noindent
{\bf All sky camera.}
\noindent
An all-sky camera KAI 340 CCD (640$\times$480 px) with a 1/2'' c-mount fisheye lens of $f/1,4-16$ has been installed recently to measure the brightness of the night sky and detect clouds. The software for analysing its data and include them in the MAGIC analysis is currently under development. The idea is to complement the LIDAR information in the sense of identifying those next target regions of the sky, where the presence of clouds would discourage data taking. 
This information may be used to change the source schedule dynamically during the night. An example of an obtained image is given in Fig.~\ref{fig2}. \\

\noindent
{\bf Atmoscope.}
\noindent
In May 2012 an ATMOSCOPE unit was installed at the MAGIC site in La Palma. ATMOSCOPE is the standard atmospheric monitoring station of the CTA collaboration~\cite{Bulik,Gaug}  and includes a Light of Night Sky (LoNS) sensor based on a HAMAMATSU 3584 PIN-photodiode and a filter wheel, and a radiation thermometer as cloud altitude sensor (Fig.~\ref{fig1}).\\


\noindent
{\bf ATCA website.}
\noindent
The ATCA website
is currently structured in two different subsites. First, there is a standard WS webpage that includes all WS data together with Meteosat movies and alerts for operators. Apart from the current values, updated every 15\,s, there are displays of historical weather data of temperature and dew point, relative humidity, wind average and gusts, and wind direction, all available for the last 3, 6, 12 and 24 hours (Fig.~\ref{fig3} top). The second website is dedicated to show the information on clouds and dust. It includes Eumetsat movies for medium and high altitude clouds, the dust content measured at the Telescopio Nazionale Galileo and the last LIDAR measurement with the information derived from it; i.e., the aerosol transmission integrated at 4 different heights (3, 6, 9 and 12 km) and the height of the lowest cloud (Fig.~\ref{fig3} bottom). The all-sky camera images and the LoNS obtained with the atmoscope will be integrated in this website together with the cloudiness reported by the pyrometer.

A third website is planned for the future, in which high level information obtained from all ATCA instrumentation will be displayed. The main goals of the analysis of ATCA instrumentation data are i) the determination of the 2-4 most common scenarios and the classification of current atmospheric conditions in one of them, and ii) the evaluation of their impact on the telescope performance. This information will be displayed by showing the on-line energy threshold: a parameter that depends much on atmospheric conditions and would allow to adapt dynamically the schedule according to the source characteristics. The correction factors to be applied to obtain unbiased reconstructed energy and flux, the identification of current atmospheric conditions, and the regions of the sky that present high cloudiness will be shown as well.\\


\vspace{-0.7cm}
\section{MC studies}




The effects of aerosol layers on $\gamma$-ray shower images have been investigated with MC simulations, including different types of aerosol layers: enhancements of the ground layer, as sometimes found after calima events~\cite{Dorner,lombardi}, cloud layers at different heights~\cite{Fruck} and with different particle densities, and a high layer of aerosols, found after strong eruptions of stratovolcanoes~\cite{garciagil}. Details of these simulations and on their analysis are presented in~\cite{garrido}.

\begin{figure}[h!t]
\centering
\includegraphics[width=0.35\textwidth]{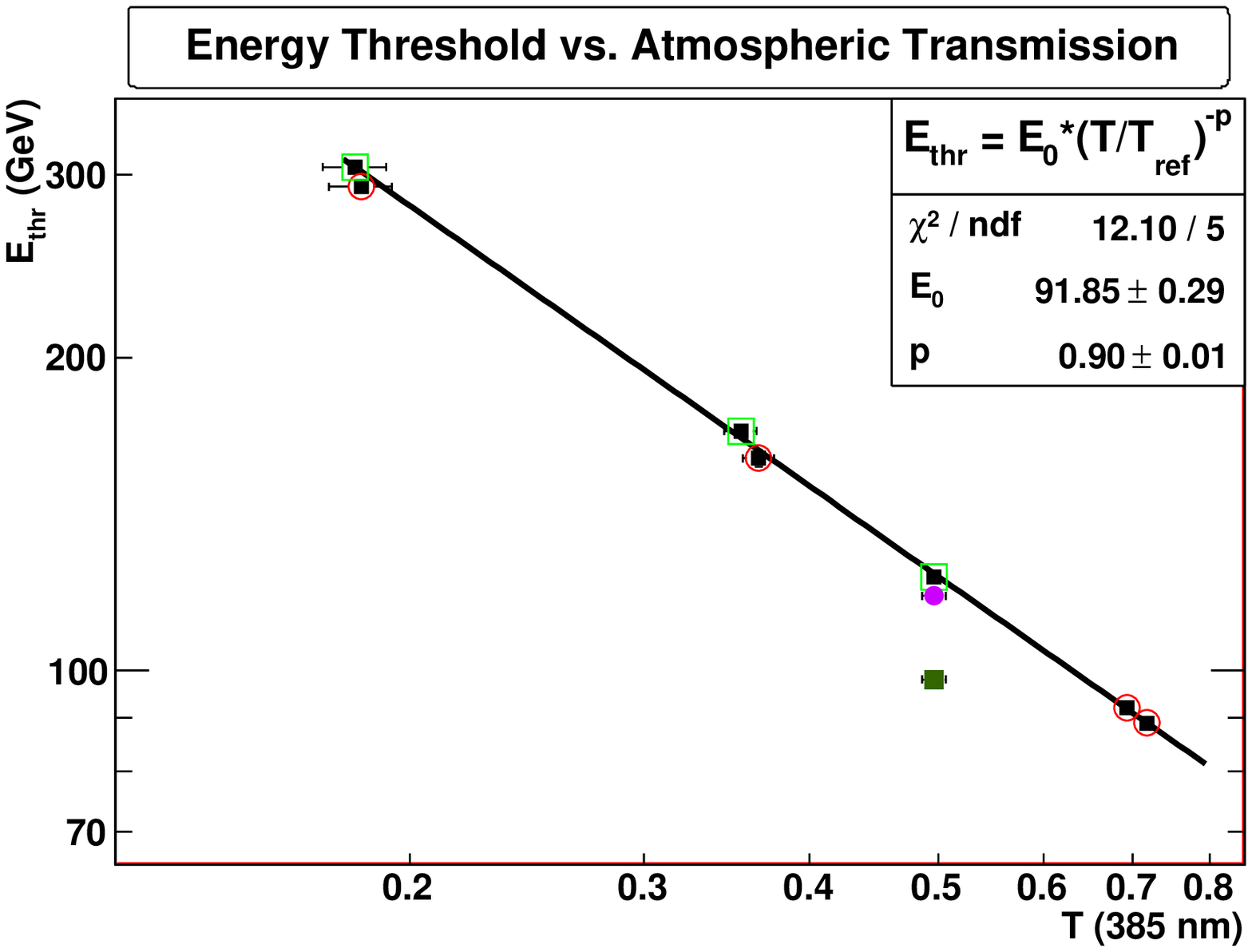} 
\includegraphics[width=0.35\textwidth]{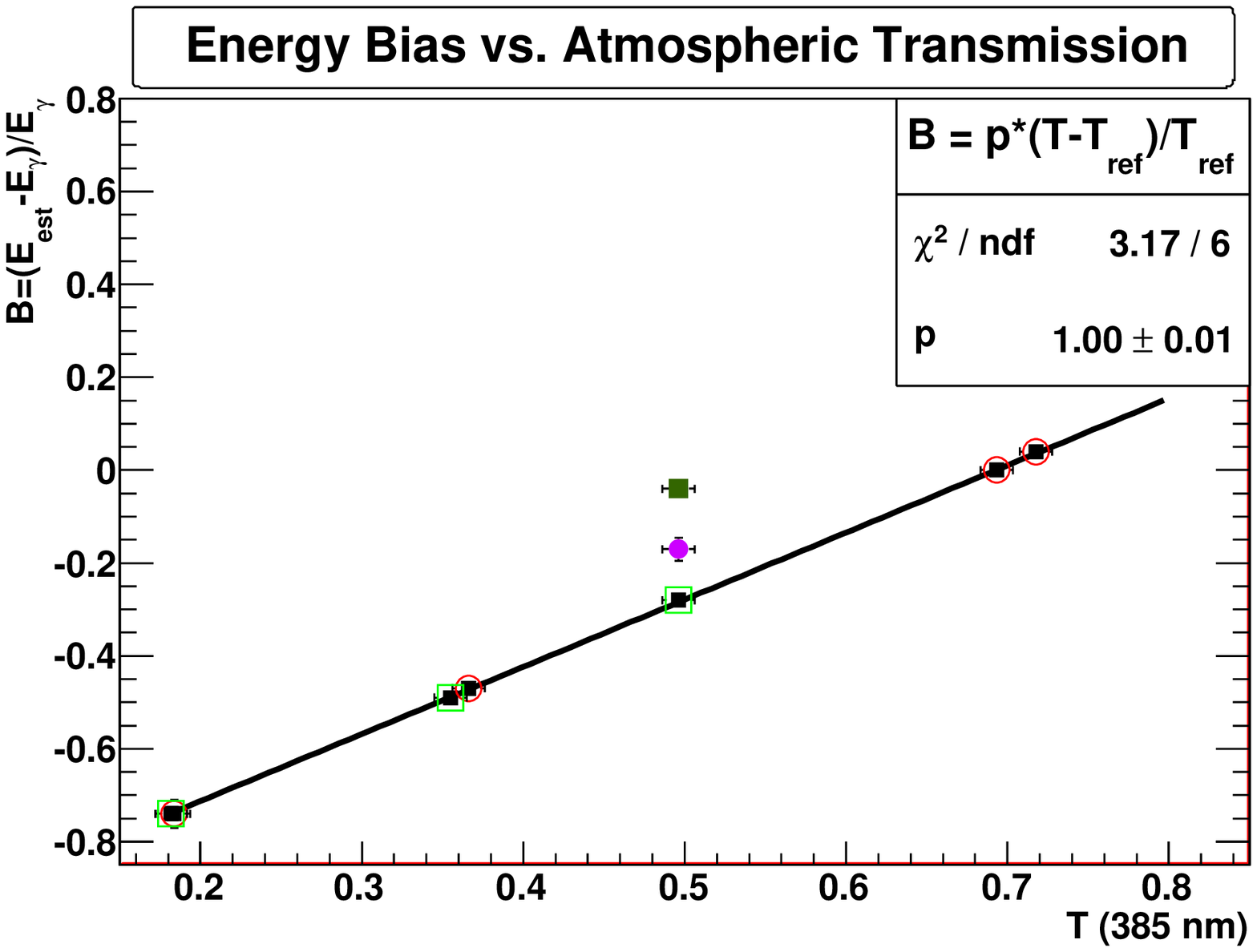} 
\includegraphics[width=0.35\textwidth]{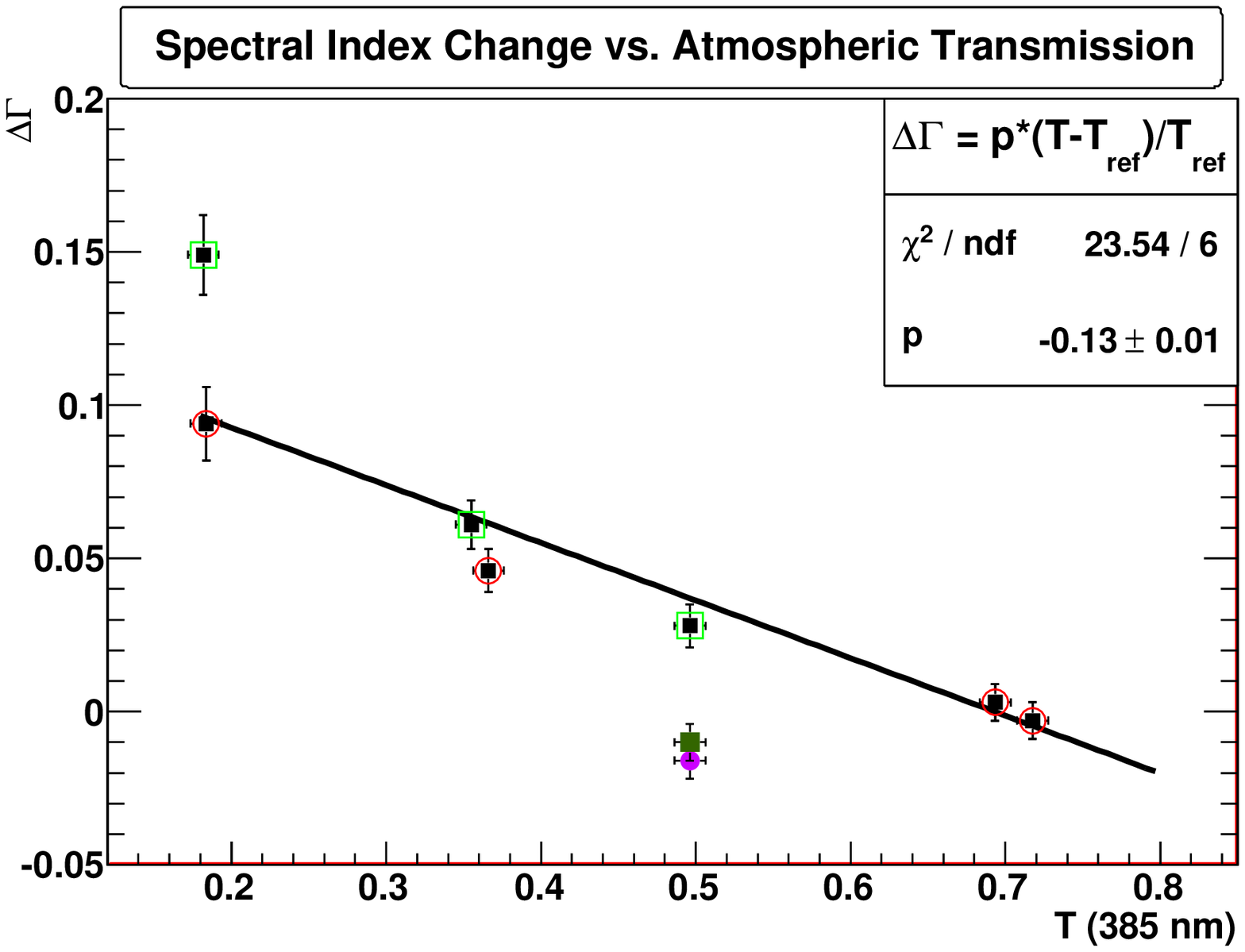} 
\vspace{-0.4cm}%
\caption{\small{%
Top: energy threshold after cuts, center: energy reconstruction bias, bottom: change of reconstructed power-law spectral index, 
all as a function of atmospheric transmission, 
measured at the wavelength of the Cherenkov peak, folded with the acceptance curve of the photo-detector.  
Open red circles: different simulated ground layer aerosol concentrations, open green squares: a simulated cloud with different particle densities, located at 
6~km a.s.l., pink point: a simulated cloud at 10~km a.s.l., dark green filled square: a stratospheric layer at 14~km a.s.l.
\label{thr}}\vspace{-0.4cm}} 
\end{figure}

We found that the effect of low clouds is very similar to that of ground layer enhancements. This can be seen in Fig.~\ref{thr}, where three exemplary parameters are shown as a function of integral atmospheric transmission: the energy threshold after cuts, the energy reconstruction bias, and the change of reconstructed spectral power-law index, always if (wrong) clear atmosphere reconstruction tables are used for the energy and flux reconstruction. All three parameters scale roughly as the inverse of, or linearly with, the atmospheric transmission, independently of the height of the aerosol layer, until about 6~km a.s.l. At higher cloud altitudes the scaling law does not apply any more. 

Corrections of energy and flux are hence rather straight-forward, i.e. simple scaling factors can be used, if the aerosol 
layer is found at low altitudes. This has already been found earlier~\cite{Dorner}. In the case of layers at higher altitudes, more sophisticated methods need to 
be developed, such as the method presented in~\cite{Fruck}, or even a dedicated inclusion of the layer in the construction of the likelihood, when a model analysis 
is performed~\cite{modelanalysis}. 

\vspace{-3mm}
\section{Determination of atmospheric conditions.}

To date, the standard MAGIC MC simulations consider two different atmospheric molecular profiles, denoted by MAGIC winter and MAGIC summer, and a single aerosol profile, known as Elterman model~\cite{Elterman} which corresponds to the standard U.S. clear atmosphere. The MAGIC summer and winter molecular profiles were derived from public radiosondes data available online~\cite{Haffke, BADC} and describe quite well the atmosphere molecular density in summer (from May to October) and winter (from November to April), being the deviations between measured and model densities lower than 2\% respectively~\cite{Biland}.

\begin{figure}[h!t]
\vspace{-0.3cm}%
\centering
\includegraphics[width=0.39\textwidth]{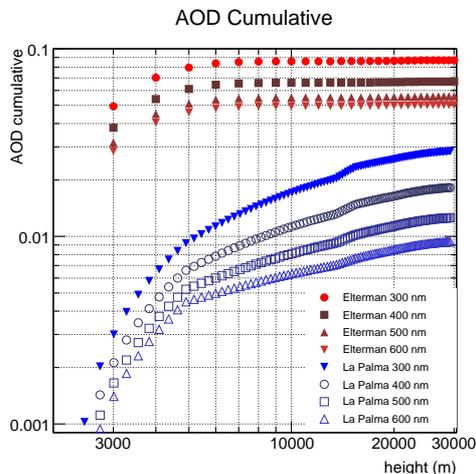} 
\vspace{-0.5cm}%
\caption{\small{%
Cumulative Aerosol Optical Depth as a function of atmospheric height, from the ORM. 
The red markers show the Elterman model for different wavelengths, the blue marker the new model for La Palma.
\label{fig5}}\vspace{-0.5cm}} 
\end{figure}

Over the last ten years, a precise picture of the aerosol composition of the free troposphere above the Canary Islands, and its seasonal variations~\cite{lombardi,garciagil,andrews,rodriguez} has been obtained. We found that the currently used aerosol model for clear nights
shows too big aerosol concentrations at low altitudes. Based on these findings, we have constructed a new aerosol model for clear nights, which is currently being tested with simulations, and cross-validated with data. Figure~\ref{fig5} shows the integral aerosol optical depth from ground to a test height, for different wavelengths and for old (Elterman) and new (La Palma) model.


Apart from that, using an almost continuous sample of one year of nightly LIDAR data, we will proceed to classify 3 typical scenarios 
for the atmosphere above La~Palma, and generate more realistic samples of MC simulations, which will be used as a test tool for present and future data correction 
algorithms. Already now, we see that the most common situation is the clear night, i.e. no aerosol layer visible in the LIDAR return. Next, a thin aerosol layer 
at about 8--10~km a.s.l. appears from time to time which does not reduce considerably the trigger rates, but requires some correction, at least for the low-energy events.
We expect for this Summer also several calima events, with which our atmospheric monitoring equipment, especially the LIDAR, can be cross-validated 
with external instrumentation at the ORM~\cite{lombardi}.

The production of dedicated MC to reproduce the different atmospheric conditions is very CPU consuming and may not be the optimum way to obtain unbiased energy and flux reconstructed unless we can identify very few common scenarios. 

An alternative way
is to use the cumulative transmission profile obtained with the LIDAR~\cite{Fruck}. It basically consists of correcting the image size the factor by which the light content of the air-shower image has been reduced, and evaluating the effective collection area at the energy before up-scalling. This method is based on the assumption that an air-shower affected by aerosol extinction can be treated like an air-shower from a lower energy primary particle, regarding trigger efficiency and energy reconstruction for low to moderate aerosol extinction.

\vspace{-3mm}
\section{Correlation studies and real data analysis.}

\begin{figure}[h!t]
\vspace{-0.2cm}
\centering
\includegraphics[width=0.42\textwidth]{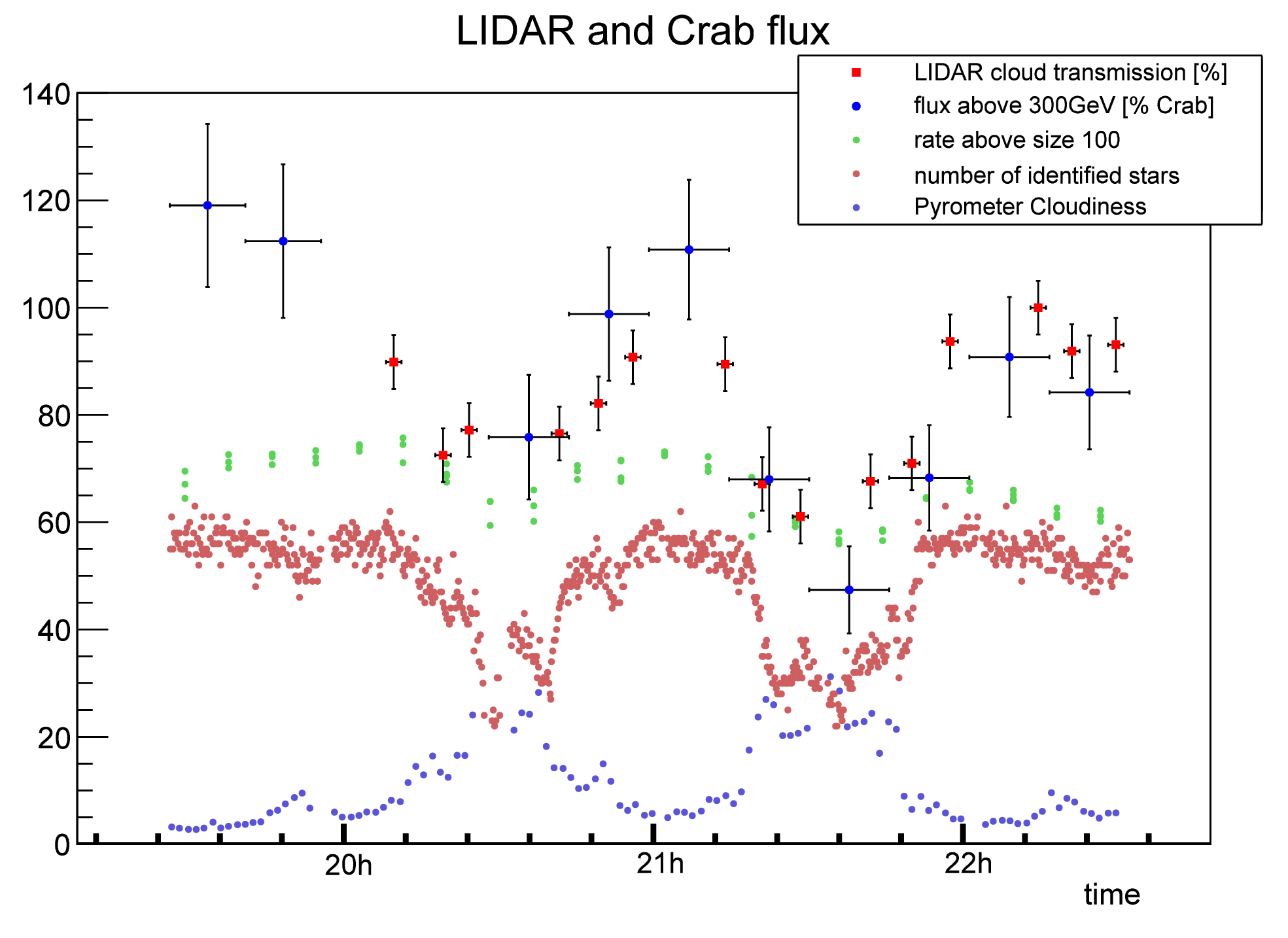} 
\vspace{-0.3cm}%
\caption{\small{%
Comparison of Crab flux above 300 GeV with LIDAR cloud transmission, trigger rates, number of identified stars and pyrometer cloudiness.
\label{fig6}}\vspace{-0.3cm}} 
\end{figure}

ATCA instrumentation are providing data that could be used to determine the atmospheric conditions above the telescopes. We are performing correlation studies between ATCA data with two main goals. First, such studies will generate understanding on the atmosphere behavior and could be used to cross-check the different instruments, making easier the detection of instrument malfunctioning, for instance. Second, if good correlations are found, we could try to recover part of the data taken under low or moderate cloudy conditions that have not been considered in the former analysis so far. Fig.~\ref{fig6} shows an example of possible correlations and anticorrelations that can be found by comparing the Crab flux above 300 GeV with different ATCA instrumentation data.

\vspace{-3mm}
\section{Conclusions and perspectives.}

Different instrumentation has been set up at the MAGIC site to determine the atmospheric conditions.
MC studies have shown the need of measuring range-resolved atmospheric transmission and how correction factors can be applied to the reconstructed energy and flux with the price of increasing the energy threshold. In conditions when Mie scattering is dominated by an aerosol ground
layer, a linear relationship between the energy threshold and the inverse of the atmospheric transmission has been found, as well as between the energy bias and the atmosphere transmission. In addition, a new LIDAR-based atmospheric calibration method has been developed to correct the same parameters (energy threshold and bias) by correcting the event size from the atmospheric transmission profile. These two approaches will constitute two independent ways to finally obtain on-line the energy threshold of our telescope during data taking. This information, together with the regions of the sky with high cloud coverage
will constitute the main information that could be potentially used by a smart source scheduler. The analysis of the atmospheric instrumentation data set together with the correlation studies will hopefully allow the MAGIC collaboration to increase the duty cycle of the telescopes by a factor of up to 15\%.

\bibliographystyle{plain}

\end{document}